\documentclass[pra,twocolumn,showpacs,amsmath,amssymb]{revtex4} 

\usepackage{graphicx}       
\usepackage{txfonts}        
\DeclareMathAlphabet{\mathcal}{OMS}{cmsy}{m}{n}	

\newcommand{\gvec}[1]{\boldsymbol{\mathrm{#1}}}			

\begin{document}


\title{Vortex-lattice formation and melting in a nonrotating Bose--Einstein condensate}

\author{Gary Ruben}
\email{gary.ruben@sci.monash.edu.au}
\author{David M. Paganin}
\author{Michael J. Morgan}
\affiliation{School of Physics, Monash University, Victoria 3800, Australia}

\date{June 27, 2008}

\begin{abstract}

Numerical simulations of the interference of a three-way segmented nonrotating Bose--Einstein condensate reveal the production of a honeycomb vortex lattice containing significant numbers of vortices and antivortices. If confined within a trap, the lattice subsequently melts, exhibiting a rich assortment of vortex--antivortex interactions. In contrast with nonlinear vortex production mechanisms previously described for Bose--Einstein condensates, the process here is shown to be primarily one of linear superposition, with initial vortex locations approximately described by a linear theory of wave packet interference.

\end{abstract}



\maketitle

\section{Introduction}

The production of vortices has attracted great interest in the study of Bose--Einstein condensates (BECs) (see, e.g., \citep{Pis99, DalGi99, FetSv01, MinSu04}). Typically, production has been through bulk rotation of the condensate cloud, such as with a ``laser spoon'' or by laser phase imprinting. These rotating systems form an Abrikosov lattice \citep{Abr57} of vortices with hexagonal symmetry. In contrast with the rotating BEC, in which the number of vortices is governed by the net angular momentum of the system, the nonrotating BEC can also give rise to vortices due to the reconciliation of initial random phase variations via the Kibble--Zurek mechanism \citep{AngZu99}.

Interference of two nonrotating BEC pieces with a repulsive nonlinearity, has also been shown to give rise to vortices \citep{FedPi00}. In this system, analogous to the Young's two-pinhole interferometer, the interference fringes---also known as dark stripe solitons---decay via a ``snake instability'' into a string of vortices \citep{KivLu98,FedPi00,BraRe02}. This vortex formation mechanism relies on the nonlinearity of the BEC self interaction.

\begin{figure}[bt]
    \centering
    \includegraphics[width=74mm]{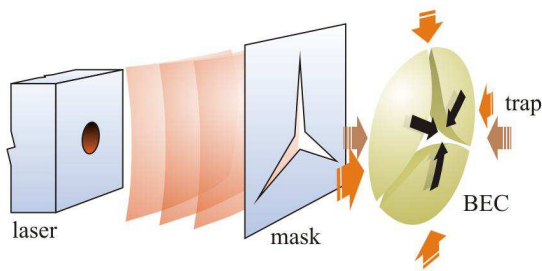}
  \caption{(Color online) A laser-illuminated mask separates a pancake-shaped BEC, formed in an asymmetric trap, into three pieces. Upon removal of the illumination---and optionally the trap---the pieces interfere, forming a lattice of vortices and antivortices.}
  \label{fig:schematic}
\end{figure}

Recently
\citet{SchWe07}
performed an experiment in which vortices
were observed as a result of the interference of a three-way segmented BEC. An oblate spheroidal BEC was formed in an asymmetric trap partitioned into three regions by shining a laser light sheet on the condensate (see Fig.~\ref{fig:schematic}).
The partition walls were then removed at varying rates and vortices were sought after different elapsed times, during which the three pieces were allowed to expand and interfere. Both 2D and 3D numerical simulations of this experiment have also been performed by \citet{CarAn08}.

In this paper, we demonstrate with numerical simulations that vortices are produced by a three-segment BEC devoid of initial phase variations and show that this mechanism is, in contrast with the two-piece case, predicted by a linear theory, the development of which is related to previous work on the three-pinhole Young's interferometer
\citep{RubPa07}.
In contrast with \citet{CarAn08}, we have not sought to replicate the aforementioned experiment exactly. By instead focusing attention on those initial states characterized by a constant phase, we aim to demonstrate that initial phase variations are unnecessary for vortex formation. In addition, we demonstrate interference and vortex production in the absence of a confining transverse trap, thereby reducing the requirement for the nonlinear processes at play in two-fragment condensate interference, and providing deeper insight into the argument that the vortex generation mechanism for three symmetrically arranged, well-separated pieces is, by contrast, primarily a linear process.

By increasing the intensity of the light sheet, we have also been able to generate a lattice comprised of significant numbers of vortices and antivortices. In the trapped system, the regular vortex--antivortex lattice subsequently melts, exhibiting a diversity of interactions by, for example, self-propelling vortex--antivortex dipoles (VDs) \citep{NozPi90, CraVe03, MVi05, SakHi06, 
PieM06,
KleJa07, CarAn08}, rotating vortex tripoles and quadrupoles \citep{MVi05, PieM06}. These interactions between vortices and vortex clusters in the condensate include dipole scattering and annihilation events. The large vortex population produced in the trapped system makes it an excellent environment for the study of vortex dynamics, VDs and vortex--sound interactions \citep{Pis99}.

The remainder of this paper is structured as follows. We describe the production of vortices via the phasor approach in Sec.~\ref{sec:PhasorDescription}. In Sec.~\ref{sec:NumericalModel} we describe the numerical BEC model, focusing on the time-dependent and ground state models, the reduction from 3D to 2D, and a vorticity metric for the order parameter field. We present the simulation results in Sec.~\ref{sec:Results}, highlighting the effects of light-sheet intensity, different light sheet geometries and phase variation between the condensate pieces in the trapped and untrapped cases, whilst making connections to the linear theory. We provide a visualization of the lattice formation and melting, and in Sec.~\ref{sec:VortexDynamics}, describe the rich vortex--antivortex dynamics that arise after the lattice has melted. Finally, we summarize our findings in Sec.~\ref{sec:Conclusion}.


\section{\label{sec:PhasorDescription}Phasor Description of Vortex Production}

In weakly nonlinear systems, typical of experimentally realistic BEC systems, we expect to observe phenomena approximately predicted by the linear theory. The creation of a vortex--antivortex lattice is one such observation, whose geometry is described in the Appendix. We include this short section to explain the production of vortices by linear interference, from which the lattice is formed.

\begin{figure}[bt]
    \centering
    \includegraphics[width=85mm]{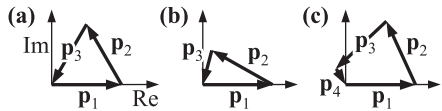}
  \caption{Representing the sources with a phasor diagram, vortices lie at locations corresponding to a closed loop of phasors. (a) Three sources linearly superposed. (b) Two sources and a nonlinear interaction term. (c) Four sources linearly superposed or three sources plus a nonlinear interaction term.}
  \label{fig:phasors}
\end{figure}

In order to form vortices by a linear interference process, at least three source waves are required \citep{MasDu01}. This may be understood with reference to the phasor diagram Fig.~\ref{fig:phasors}(a) in which each phasor represents a source wave.
The probability amplitude is zero at a vortex core, corresponding to a closed loop comprised in general of three or more phasors. Although two phasors may sum to zero, such a case corresponds to a domain-wall defect rather than a vortex. Vortex formation from linear superposition of higher numbers of sources has also been demonstrated \citep{OHoPa06}, corresponding to a loop of four or more phasors.

In the following, the individual terms of a sum of complex scalar waves $\Psi_1+\Psi_2+\cdots$ are represented graphically as phasors $\gvec{p}_1+\gvec{p}_2+\cdots$ in Fig.~\ref{fig:phasors}.
Figure~\ref{fig:phasors}(a) is a representation of the linear interference of three sources $\Psi_1+\Psi_2+\Psi_3$ evaluated at the vortex location. Alternatively, Fig.~\ref{fig:phasors}(b) represents a nonlinear system in which the third contribution is now associated with the interaction term in the sum $\Psi_1+\Psi_2+\mathrm{f}(\Psi_1,\Psi_2)$. The production of vortices by the snake instability might be represented by the latter picture. Thus, a loop of $n$ phasors may correspond either to the interference of $n$ sources in a linear system, or to $n-1$ sources plus a nonlinear interaction term. For example, Fig.~\ref{fig:phasors}(c) may represent either four sources $\Psi_1+\Psi_2+\Psi_3+\Psi_4$ in the case of a linear system, or three sources plus a nonlinear interaction term $\Psi_1+\Psi_2+\Psi_3+\mathrm{f}(\Psi_1,\Psi_2,\Psi_3)$ in the case of a nonlinear system.

Recently, the results of the interference of 2D condensates sliced into two and four pieces have been studied \citep{CarWh08}. Examination of these results for the case of four pieces suggest that sufficient contributions from at least three of the four source pieces have combined in the manner just described, which we refer to as a primarily linear process. For the system we study in this paper, we present further evidence for the role of the linear interference mechanism in Sec.~\ref{sec:Results}.


\section{\label{sec:NumericalModel}Numerical Model}

In this section we describe the derivation and parameters of our numerical model for the interference of a three-way segmented nonrotating BEC. We model an experiment similar to that
conducted by
\citet{SchWe07}.

We note the following features that distinguish our simulation from the experiment and its modeling by \citet{CarAn08}. First, in establishing the initial condition prior to time evolution we obtain the ground state with a uniform global phase, so that no initial phase variation is allowed within or between the three pieces, in order to demonstrate vortex production as predicted by the linear theory.
In contrast, \citet{CarAn08} apply phase variations within their numerical scheme to obtain their initial condition.
Although we subsequently also apply an example of different relative phases to the three pieces following establishment of the ground state, this is presented to demonstrate how the linear theory applies to this case.
The second distinguishing feature is the instantaneous removal of the light sheet at $t=0$. In contrast,
\citet{SchWe07} remove the light sheet at varying finite rates, but report a maximum
efficiency of vortex generation for the maximum observable rate of removal. Thirdly, we exclusively model the system in 2D. As such, our simulation results are most pertinent to pancake-shaped condensate clouds, in which the ratio of the axial to transverse trap frequencies is large in the sense outlined below. Because non-pancake condensates admit the existence of vortex lines with more complex geometries, investigation of these systems would have to be performed separately to assess the applicability of the results reported here.
A final difference is that we present simulations with and without the transverse trap to demonstrate that lack of transverse confinement is not an impediment to vortex production, consistent with predictions from the linear theory.

The production of vortices in pancake-shaped condensates may be numerically modeled in two rather than three spatial dimensions. In this case the evolution of the macroscopic wavefunction or order parameter $\Psi$ of the BEC is considered to be governed by the Gross-Pitaevskii equation (GPE) in (2+1)D,
\begin{equation}
	\label{eqn:GP}
	i \hbar \partial_t \Psi = \left[ -\frac{\hbar^2}{2m}\nabla_\perp^2 + V(r,\theta) + \gamma N U_0 \left| \Psi \right|^2 \right] \Psi,
\end{equation}
where we have adopted plane-polar coordinates for space $(r, \theta)$ and $t$ denotes time. The GPE takes the form of the (2+1)D time-dependent Schr\"odinger equation with an additional nonlinear self-interaction term $\gamma NU_0 \left| \Psi \right|^2 \Psi$ for a condensate containing $N$ atoms, each of mass $m$. Here $U_0=4\pi \hbar^2 a/m$ depends on the \textit{s}-wave scattering length $a$, which in this case is positive in order to permit the BEC pieces to expand and interfere. The multiplication factor $\gamma$ arises as a result of reduction from the full 3D description to 2D in the manner described below. In this form of the GPE, the normalization condition is $2\pi \int^\infty_0 |\Psi(r,\theta)|^2 \, r\, dr = 1$ (see, e.g., \citep{HolCo96, LeeMo02, BaoJa03}).

The axial dimension of the BEC is determined by the axial trap component $m\omega_z^2 z^2/2$, whose angular frequency $\omega_z$ provides a characteristic length $a_z=[\hbar/(m \omega_z)]^{1/2}$. When $a_z$ is much greater than the scattering length $a$, the multiplication factor $\gamma$ may be found by separating the 3D order parameter into a product of independent axial and transverse components $\Psi_{3D}(r, \theta, z)\approx\Phi_0(z) \Psi(r, \theta)$ and evaluating the expectation value of the transverse component $\Psi(r, \theta)$ along the axial direction $z$ \citep{TanMi02, LeeMo02}. The axial condensate profile is assumed to be the well-known ground-state solution $\Phi_0(z) = (\pi a_z^2)^{-1/4} \exp[-z^2/(2 a_z^2)]$ to the 1D time independent Schr\"odinger equation. The 2D Laplacian operator arises from separation of the full 3D Laplacian into its transverse and axial components, $\nabla^2\equiv\nabla_\perp^2 + \partial_z^2$. The final result is a (2+1)D GPE with a nonlinear multiplication factor $\gamma=[m\omega_z/(2\pi\hbar)]^{1/2}$ that represents the averaged value of the 3D nonlinear factor over all $z$. An alternative approach, and that presented here, is to directly evaluate the nonlinearity occurring at $z=0$, corresponding to the location of the peak density of the axial profile, and assign this value to $\gamma$. Since the nonlinearity achieves a maximum in this plane, it arguably represents a good choice for modeling the vortex dynamics. In this case $\gamma=|\Phi_0(0)|^2=[m\omega_z/(\pi\hbar)]^{1/2}$.
We have performed simulations with both approaches and find no qualitative difference with any of the effects reported here. The reader preferring the former approach may note that the factor of $\sqrt{2}$ by which the $\gamma$ factors differ may be absorbed as a change in the number of condensate atoms $N$.

The BEC ground state is determined in the presence of a trapping potential $V(r,\theta)$ formed from the combination of a harmonic trap and light-sheet potential $V(r,\theta) = m\omega^2 r^2/2 + L(r,\theta)$, where $\omega$ is the angular frequency of the radial trap component. By assuming that the light sheet $L(r,\theta)$ is instantaneously removed at $t=0$, the light-sheet potential term appears only in the time independent GPE used to determine the ground state. The harmonic trap term is included in the time dependent GPE for the trapped condensate simulations, but it is removed for the untrapped simulations.
The parameter values used in our simulations were $a=5.77\times 10^{-9}~\mathrm{m}$ for \textsuperscript{87}Rb \citep{DalGi99}, $N=2.6\times 10^5, \omega=46.5~\mathrm{rad~s}^{-1}$ and $\omega_z=88.6~\mathrm{rad~s}^{-1}$.

We define a vorticity metric for the order parameter field $\Psi$ according to 
\begin{equation}
	\label{eqn:vorticity}
	\mathrm{vorticity} = \dfrac{1}{V}\iint \big| \nabla \! \times \gvec{j} \big| \ dx \ dy,
\end{equation}
where the integral represents integration over the numerical field of volume $V$, and $\gvec{j}$ is the probability current
$\gvec{j}\equiv (\Psi^* \nabla \Psi-\Psi \nabla \Psi^*)\hbar/(2m)$.

This metric is similar to the usual vorticity measure in fluids $\gvec{\omega}=\nabla \times \gvec{v}$, of the velocity field $\gvec{v}=\gvec{j}/\rho$ where $\rho=|\Psi|^2$ is the local probability density. By not dividing the probability current by $\rho$, extra weighting is conferred upon vortices located within a locally increased density over vortices in regions of lower density. The metric sums the modulus of the local measure over the whole field.

\section{\label{sec:Results}Results}

\begin{figure*}[tb]
    \centering
    \includegraphics[width=174mm]{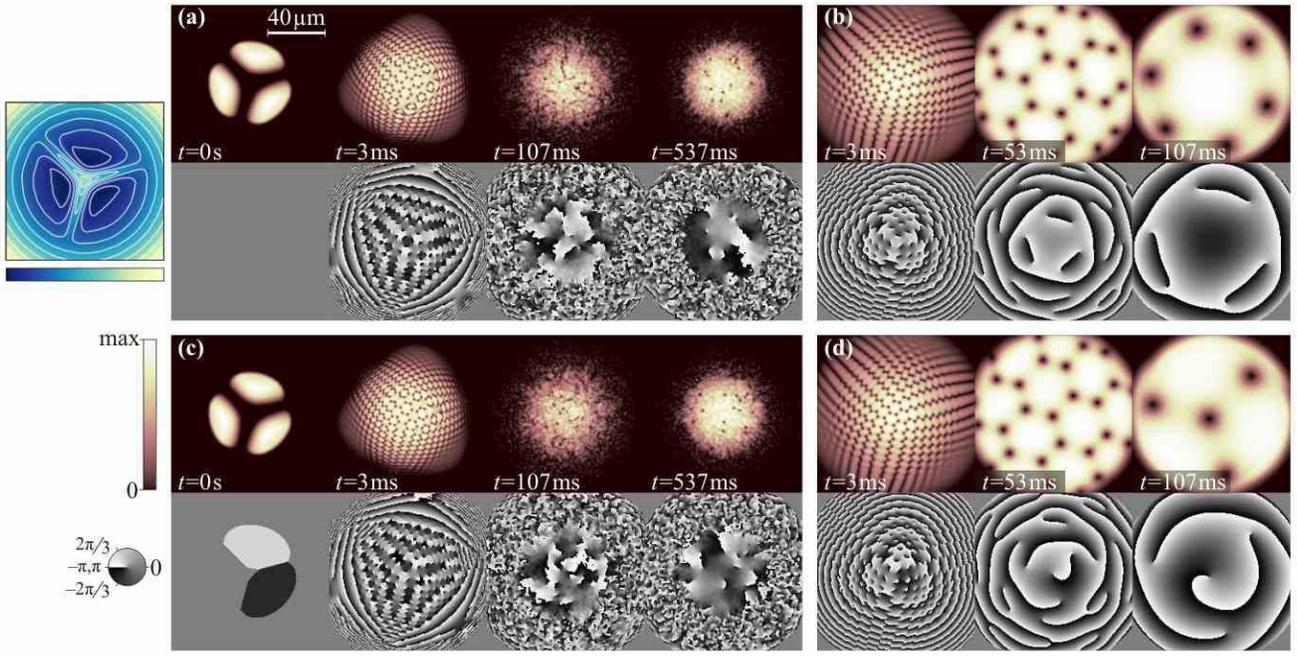}
  \caption{(Color online) Amplitude (upper) and associated phase plots (lower) for the ground-state trap potential shown as a contour-plot at left, which is a 2D harmonic trap with a superposed three-way light sheet of dimensionless amplitude $I_0=0.8$. The trap profiles identifying wall height $h_0$ and width $w_0$ are taken along the dotted paths. (a) Three-segment BEC interference with a harmonic confining trap showing progression from the ground state, through lattice formation to a late-stage characterized by complex vortex--antivortex (VA) dynamics. (b) Honeycomb lattice formation and free expansion in the absence of a confining trap. (c-d) The initial piece phases are rotated to 0, $2\pi/3$ and $-2\pi/3$. The effect is to shift the lattice so that an antivortex coincides with the trap center.}
  \label{fig:results_set1}
\end{figure*}

In Fig.~\ref{fig:results_set1}
we present simulation snapshots for the segmented BEC showing the initial state, the vortex--antivortex lattice that forms after the three pieces have interfered, and later stages that, in the trapped cases, follow the disintegration of the lattice. For corresponding movies, see \citep{Movies}.
The intensity of the laser light sheet relative to the trap is represented by a dimensionless quantity $I_0$, where the trap potential contour plots (shown to the left of the time series) indicate the magnitude relative to the trap. We have chosen to present results for $I_0=0.3$ and $I_0=0.8$, as these values allow exposition of the different behaviors of poorly and well separated BEC pieces, respectively. The walls are characterized, as shown in the trap profile plots, by wall height $h_0$, expressed in units of energy, and full width at half maximum width $w_0$. $I_0=0.3$ and 0.8 correspond to $h_0\approx ~k_B \times 70~\mathrm{nK}$ and $k_B \times 180~\mathrm{nK}$, respectively; both higher than the $k_B \times 26~\mathrm{nK}$ barrier produced by \citet{SchWe07}. Our wall width $w_0\approx 9~\mathrm{\mu m}$ may be challenging to produce experimentally. A comparison value is not given in \citet{SchWe07}, although the condensate profiles are indicative of wider walls.
The upper time series show the evolution of the BEC beginning from the global ground state. Probability density and phase are also shown in Fig.~\ref{fig:results_set1}. The expansion and interference occurs both within a confining harmonic trap [Figs.~\ref{fig:results_set1}(a) and (c)] following instantaneous removal of the light sheet, and in the absence of a transverse confining trap [Figs.~\ref{fig:results_set1}(b) and (d)]. In the latter case, the condensate would expand beyond a finite simulation region. To allow for this in the numerical model, a damping term is added to the time evolution equation to absorb the outward propagating matter. Since these are 2D simulations, and as such are applicable to pancake-shaped condensates, experimental realization would most likely require maintenance of the axial trap component in both cases, not just to allow for establishment of the lattice, but also for any free-expansion stage prior to imaging.

\Citet{SchWe07}
observed vortices consistent with production by the Kibble-Zurek mechanism. They reported that 10\% of nonsegmented condensates contain vortices.  In our simulations, by starting from a ground state, no vortices may be produced by this mechanism.

\begin{figure}[tb]
    \centering
    \includegraphics[width=85mm]{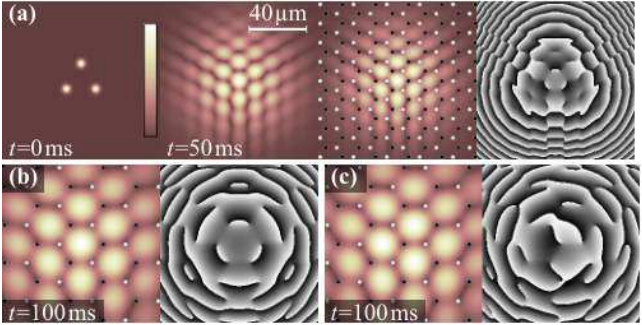}
  \caption{(Color online) Three linearly superposed wave packets lying at the corners of an equilateral triangle [cf. Figs.~\ref{fig:results_set1}(b) and (d)]. The amplitude, with analytically determined vortices (dark) and antivortices (light) overlaid [see Eqs.~\eqref{eqn:theta_mn} and \eqref{eqn:r_mn}], and phase of Eq.~\eqref{eqn:general_sum} are shown. (a) Equal-phase wave packets shown at 0~ms, after interference at 50~ms (with and without the overlaid lattice), and (b) after 100~ms. (c) Phases 0, $2\pi/3$ and $-2\pi/3$ cause a lattice translation.}
  \label{fig:linear_results}
\end{figure}

Previous work \citep{RubPa07} has shown that the linear interference of three expanding monochromatic spherical waves generates a distorted honeycomb vortex--antivortex (VA) lattice. In the Appendix, we present a related linear theory for the case of three Gaussian wave packets, evolving in (2+1)D, clarifying the effect of source phase variation on the predicted vortex locations. In this case, an infinite, regular honeycomb VA lattice is formed, with a Gaussian probability density envelope. The formation of vortices by a three-piece BEC may be understood as arising from the same mechanism, albeit now in a highly nonlinear system.
The linear theory applies most directly to the untrapped system. For observing interaction dynamics, the trap presence must be maintained. However, if the interference is instead performed with the transverse trap switched off, experimental measurement of the position of any central vortex with respect to the center and lattice parameter should allow determination of the phases of the initial BEC pieces to within a global phase factor. Figures~\ref{fig:results_set1}(b) and (d) show the formation of an extremely regular lattice, which compares favorably with that formed by linear superposition in Fig.~\ref{fig:linear_results}.

\begin{figure*}[tb]
    \centering
    \includegraphics[width=176mm]{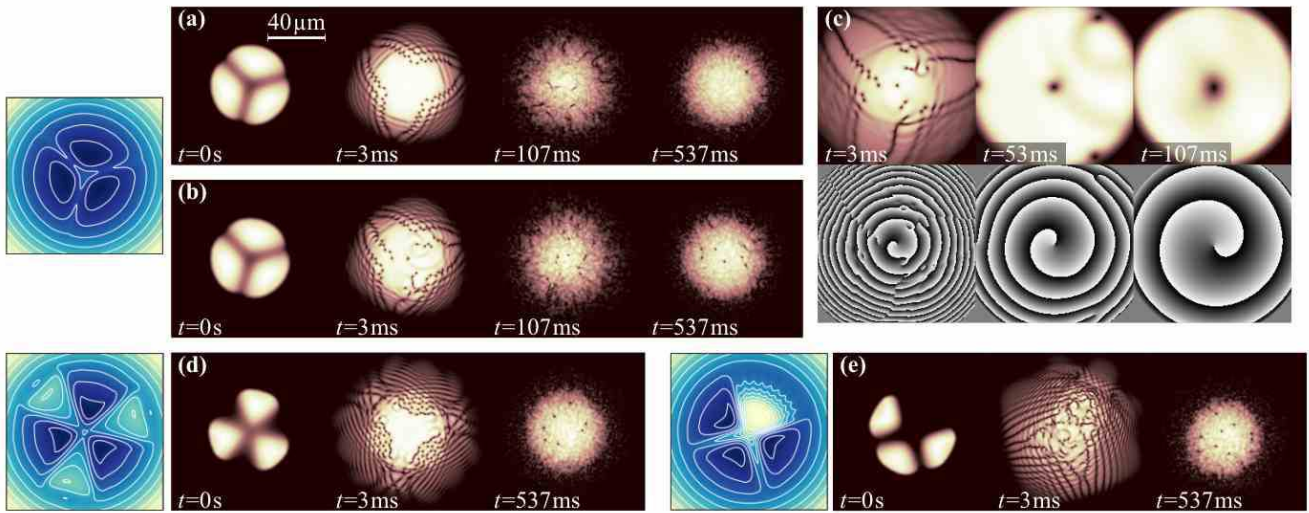}
  \caption{(Color online) Amplitude [and associated phase plots for (c)] for light sheets characterized by the trapping potentials shown as adjacent contour plots. (a-c) When compared with Fig.~\ref{fig:results_set1}, the lower intensity light sheet ($I_0=0.3$) allows vortices to propagate outwards with the condensate matter. In (a), the central region of the lattice is left devoid of vortices. In the trapped (b) and untrapped (c) cases---in which the relative phases of the initial pieces are set to 0, $2\pi/3$ and $-2\pi/3$---the central vortex is clearly visible. (d-e) Two alternative arrangements with (d) propeller-shaped pieces, and (e) pieces at three corners of a square are discussed in the main text.}
  \label{fig:results_set2}
\end{figure*}
If the three BEC pieces are sufficiently isolated from each other, they may acquire random relative phases \citep{RoeNa97}. \Citet{SchWe07} describe the presence or absence of a central vortex according to reconciliation of these phases. The linear theory instead describes an equivalent effect, which predicts the presence of a central vortex as resulting from translation of the lattice as a whole. \citet{RoeNa97} state that the pieces are ``virtually degenerate'', allowing them to be treated as coherent pieces whose relative phases may vary randomly. For a lower intensity light sheet, tunneling through the light-sheet walls ensures that the phases of the separate pieces remain coupled. The interference pattern is said to become ``locked''. In our case, the lattice translation becomes locked in an equivalent sense to give the result in Figs.~\ref{fig:results_set1}(a) and (b). To simulate the effect of decoupling between the pieces, we apply global phase factors by rotating the phase of the BEC pieces, following establishment of the ground state but prior to time evolution.

In Figs.~\ref{fig:results_set1}(c-d), the relative phases of two of the regions have been rotated from 0 to $2\pi/3$ and $-2\pi/3$, as shown in the $t=0$~s phase plot. The $t=53$~ms and $t=107$~ms cases without a transverse trap show a vortex at the center. This may be understood by evaluating the predicted vortex locations using the linear theory. To illustrate this, Fig.~\ref{fig:linear_results} provides linear simulation results for the equal [Figs.~\ref{fig:linear_results}(a-b)] and rotated [Fig.~\ref{fig:linear_results}(c)] phase cases for the model described in the Appendix. For these examples, the source positions $r_2$ and $r_3$, and the momentum uncertainty $\Delta p$ were determined by fitting Gaussian profiles to the leading (innermost) edges of the probability density of the three BEC pieces at $t=0$ for the $I_0=0.8$ case, i.e. from the first frame in Fig.~\ref{fig:results_set1}(a). Because the model assumes that the BEC pieces are well described by circularly-symmetric Gaussian pieces, which is not the case here, this approach was found to be better than fitting to the whole BEC-pieces. However, because of this departure, illustrated by the poor match of the lattice scale to that of the nonlinear results, the linear model is best applied qualitatively. Although the nonlinear dynamics of the BEC do not allow the analytically predicted vortex locations to be mapped directly onto the nonlinear simulation results, the linear theory nevertheless provides useful predictions of the generation of the central vortex and the lattice symmetry.

If the phases are such that a vortex is created sufficiently close to the center of the trap, in the case where few other vortices are produced, such as when the wall heights are low [as in Figs.~\ref{fig:results_set2}(a-d)],
the vortex may migrate to take up residence in the center of the trap. Figure~\ref{fig:results_set2}(b) illustrates this behavior. Since flows associated with phase gradients cancel there, any resident vortex occupies a privileged position and is allowed to remain there until perturbed by a local change, such as might be caused by a passing vortex set in motion following melting of the lattice.

\begin{figure}[tb]
    \centering
    \includegraphics[width=85mm]{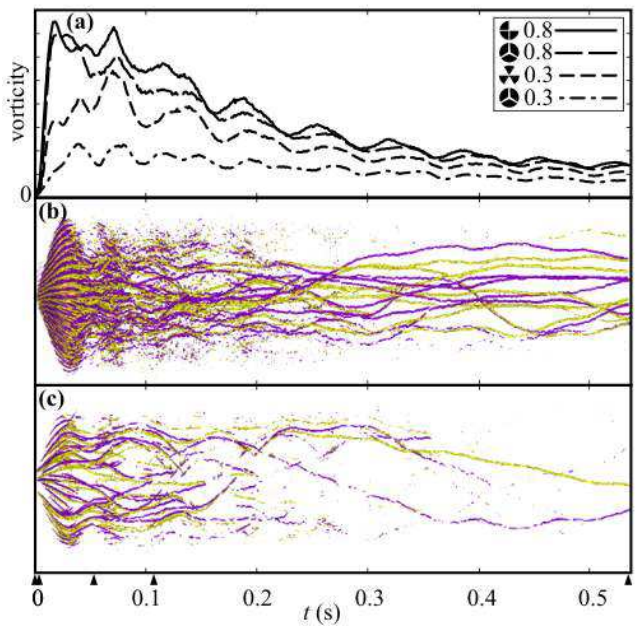}
  \caption{(Color online) Vorticity plots for selected BEC segmentation schemes with a harmonic trap. Small arrows along the time axis correspond to the times selected in Figs.~\ref{fig:results_set1} and \ref{fig:results_set2}. (a) Vorticity measure [Eq.~\eqref{eqn:vorticity}] for $I_0=0.3$ and 0.8 and for sources at three corners of an equilateral triangle, a square and propeller shaped schemes (see main text). (b) Vortices (dark) and antivortices (light) generated from the equilateral triangle scheme with $I_0=0.8$. Regions i, ii and iii are described in the main text. (c) Fewer vortices are produced with a lower intensity $I_0=0.3$ light sheet.}
  \label{fig:jellyfish}
\end{figure}

The vorticity metric is plotted against time for four different initial conditions in Fig.~\ref{fig:jellyfish}(a).
Figures~\ref{fig:jellyfish}(b-c) are time-series plots of the vortices in the side-on condensate slices. In these figures, period (i) denotes the initial formation of the regular vortex--antivortex lattice. In the subsequent period (ii), the regular lattice melts, as the condensate matter washes back towards the center due to confinement by the harmonic trap. During this period, the high population of vortices and antivortices promotes a high interaction rate characterized by a rich variety of vortex interaction dynamics. VA annihilation reduces the vortex population to a level where this rate slows and dynamics involving VDs begin to dominate in the final period (iii). The washing in and out is roughly circularly symmetric, cyclically increasing and decreasing the probability density in the center. These global oscillations are visible in Fig.~\ref{fig:jellyfish}(a) due to the vorticity measure giving extra weight to vortices embedded in a locally increased condensate probability density. The predicted frequency of the radial breathing mode of a harmonically trapped 2D condensate is twice the trap frequency $\omega$ \citep{Pit96} (In our simulations $\omega=46.5~\mathrm{rad~s}^{-1}$). The measured frequency from the lower three plot series, whose condensates have centroids coinciding with the trap center, is $93.6 \pm 0.6~\mathrm{rad~s}^{-1}$, in agreement with theory. From Figs.~\ref{fig:jellyfish}(b) and (c) we see that the vortices themselves are carried in and out by the bulk motion of the condensate matter, presumably changing the vortex interaction rate in turn. The interaction dynamics are discussed in more detail in Sec.~\ref{sec:VortexDynamics}.

A lower wall height results in the pieces starting closer together. In the linear theory in \citep{RubPa07}, correspondingly smaller $r$ values generate smaller numbers of vortices. This is clearly illustrated by the smaller vorticity values in Fig.~\ref{fig:jellyfish}(a) and the reduced population of vortices in Fig.~\ref{fig:jellyfish}(c) over Fig.~\ref{fig:jellyfish}(b). In Figs.~\ref{fig:results_set2}(a-b), we see that the matter expanding from the center carries most of the vortices to the outer parts of the trap as the lattice is forming, where they may be lost altogether, further reducing the vortex population.

The propeller shape in Fig.~\ref{fig:results_set2}(d) is included to show an alternative mask, which also produces BEC pieces at $120^\circ$ angles to each other, but which pushes the centers of the condensate pieces further apart. Accordingly, higher numbers of vortices are predicted by the BEC created from this mask. Indeed, this is the case, as can be seen by comparing the bottom two results in Fig.~\ref{fig:jellyfish}(a), which plot vorticity for the same light sheet intensity $I_0=0.3$ for the two $120^\circ$ alternatives.

In the linear theory in \citep{RubPa07}, the maximum number of vortices is predicted for a source arrangement in which the pieces are initially arranged at three corners of a square. This prediction guided investigation of a new source configuration, shown in Fig.~\ref{fig:results_set2}(e), which may be realized by a cross-shaped light sheet mask with one quadrant open to the passage of light from the illuminating laser. Confirmation of the prediction is apparent by comparing the top two results in Fig.~\ref{fig:jellyfish}(a), which are for the same light-sheet intensity. In fact, the spacing of the condensate centroids for the $90^\circ$ arrangement is smaller than for the $120^\circ$ case. Thus the increase in vorticity due to the angular arrangement more than offsets the decrease in vorticity-generating capacity caused by reducing the piece spacing. Whilst we have shown a geometry with an increased capacity to generate vortices, it is possible that nonlinearity causes the absolute maximum to be achieved for another angle close to $90^\circ$. Evidence of nonlinear effects is visible in this case: the snake instability \citep{KivLu98,BraRe02} is visible as a curvature of some fringes of the second frame in Fig.~\ref{fig:results_set2}(e). On close inspection we see that the linear vortex generation mechanism dominates in this region, as disturbances from the third BEC piece propagate through and interfere with these fringes prior to the snake instability evolving to form vortices directly. The initial displacement of the condensate centroid from the center of the trap results in the merged condensate oscillating back and forth along the initial mirror symmetry plane, in addition to the radial oscillation mode. In a previous investigation  the ability to suppress the snake instability was reported in a system containing a ring dark soliton (RDS)---a dark stripe soliton formed into a closed loop---that shared the circular symmetry of the confining harmonic trap \citep{TheFr03}. Depending on the RDS parameters, this structure could preferentially decay by emission of radiation instead of the snake instability. In the three-segment BEC system, the intervention of linear interference can be seen as another means for suppressing the onset of the snake instability in dark stripe solitons.

A possible explanation for \citet{SchWe07} not observing the production of large numbers of vortices along with a regular lattice in experiments may be their application of light-sheet intensities below our lower intensity results. Blurring and possible nonuniformity of the light-sheet walls is related to the size and manufacturing process used for the mask---attention to the design of the mask optics and optical path may permit the creation of more uniform, sharper walls, leading to the lattice structures we see in our simulations. Also, real experimental initial conditions with a non-zero-temperature chemical potential introduce some phase randomness which would disrupt the lattice. Finally (as noted in \citep{CarAn08}), if the condensate is more spheroidal than pancake-like, the probability of observing vortices in the projected 2D image is reduced due to the extra freedom of possible vortex line geometries.


\section{\label{sec:VortexDynamics}Vortex Dynamics}

In this section we focus attention on the striking interactions seen to occur between those vortices and antivortices remaining in the condensate following the disintegration of the lattice. The vortices and antivortices are propelled and evolve within a seething background sea of sound waves. Due to the complexity of these interactions, we provide a phenomenological description of just some of the rich dynamics, making observations of structures previously described by other investigators. A fuller appreciation of the complexity will be gained by viewing the accompanying movies \citep{Movies}.

The dynamics described in this section will be exhibited primarily in 2D pancake-shaped condensates. In 3D, vortices are string-like objects, either forming closed loops (e.g., ring solitons) or terminating at two points on the condensate surface \citep{MinSu04, CarCl06}. In this case, the interaction dynamics are instead characterized by string intercommutation or formation of smaller loops.

Most of the BEC vortex literature has focused on vortices (or antivortices) in rotating traps, in which antivortices (vortices) are expelled from the condensate. The complex interactions described here instead rely on the presence of a population of vortices and antivortices and hence are best observed in a nonrotating system. Interfering condensates comprised of two, three, or more pieces will all generate the required conditions, provided they are nonrotating. Another example of a nonrotating system able to generate an equal population is the ring dark soliton system \citep{TheFr03}. Experimentally, phase contrast imaging with detuned light may leave the BEC sufficiently undisturbed to allow observation of these dynamics \citep{TurDo05}.

\begin{figure}[tb]
    \centering
    \includegraphics[width=86mm]{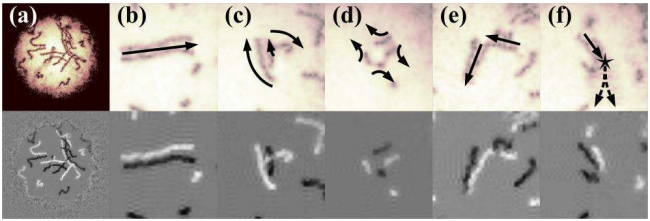}
  \caption{(Color online) Vortex interaction dynamics for the $120^\circ$, $I_0=0.8$ arrangement. In the lower frames vortices (dark) and antivortices (light) are shown. (a) Vortex trails corresponding to the range $t$=323--429~ms. (b) A propagating vortex--antivortex dipole (VD). $t$=430--471~ms. (c) Vortex--vortex--antivortex tripole rotating through $\approx90^\circ$. $t$=471--517~ms. (d) Two VDs meet to form a quadrupole. VA partners are exchanged and the new VDs move off orthogonally to the original directions. $t$=526--537~ms. (e) A VD meets a lone vortex. The antivortex exchanges its vortex partner and the new VD moves off on a new trajectory. $t$=285--328~ms. (f) VA annihilation resulting from passing by a lone vortex. $t$=342--366~ms.}
  \label{fig:paths}
\end{figure}

A vortex generates a local circulating velocity field and associated phase gradient which falls off rapidly with distance. Another vortex or antivortex in this field experiences a force in the direction of flow \citep{NozPi90}. Similarly, VDs travel at a velocity determined by their distance apart. In Fig.~\ref{fig:paths}(a),
the paths of several VDs, are shown over a 106~ms period. The dipoles in which the partners are more widely spaced have shorter paths, illustrating the predicted behavior (see also accompanying movies \citep{Movies}). In contrast, rotating vortex--vortex (or antivortex--antivortex) molecules are rarely and only fleetingly seen. Two equal-charge vortices can circulate around a common point, midway between them, like a facing pair of figure skaters. However, these structures have been shown to be unstable in radiative media \citep{Pis99, ProPa04} and would also be disturbed by the more mobile VDs, or prevented from rotating by (countering) field gradients due to the presence of nearby antivortices (vortices). 

In Fig.~\ref{fig:paths}(b), a VD approaches a lone, near-stationary antivortex, with which it couples to form the tripole shown in Fig.~\ref{fig:paths}(c). The net angular momentum of the tripole subsequently causes it to rotate through approximately $90^\circ$ before it interacts again \citep{MVi05,PieM06}. In Fig.~\ref{fig:paths}(d), a rare event is shown, in which two counter-propagating VDs approach and momentarily meet, forming a quadrupole, before exchanging partners and moving off orthogonally to the original directions. In Fig.~\ref{fig:paths}(e), a VD approaches a stationary vortex, which is swapped for the partnered vortex. The new VD continues on a new trajectory. This case may be contrasted with the formation of the tripole. In both cases, a VD approached a stationary (anti)vortex. However, the different interaction geometries resulted in the two different examples of dynamics shown.

The spacing of VDs changes in response to local field gradient perturbations, causing them to slow, speed up, separate completely, or annihilate. An example of annihilation, promoted by the proximity of another vortex, is shown in Fig.~\ref{fig:paths}(f). Following the annihilation, scattered remnant waves travel ahead of the event location, dissipating the residual VD kinetic energy. Such interactions of the vortex with field perturbations and radiation of energy as waves are examples of vortex--sound interactions \citep{Pis99}. As already mentioned, vortex--vortex molecules are unstable, emitting spiral waves as they travel apart. Indeed, the sound waves emitted by all dynamical vortex interactions travel outwards and are reflected by the trap, recombining to form a chaotic fluctuating background that affects the ongoing condensate evolution.

\citet{ParPr04} analyzed a BEC containing a lone vortex within a harmonic trap that was carefully modified to control the emission of spiral waves from a confinement region near the trap center. When spiral wave radiation was permitted to escape confinement and prevented from reinteracting with the vortex, the local energy was thus reduced, manifesting as a migration of the vortex away from the condensate center and its precession around the trap.
Although our system is complex in comparison, we nevertheless observe sound emission from isolated vortices in the form of spiral waves. As our system evolves, a reduction in the number and associated energy of the vortices also occurs, consistent with the conversion of this energy to sound.

Vortices residing in the outer parts of the trap spiral helically about the center in a right-hand screw sense. Antivortices spiral in the opposite sense and are therefore likely to meet the aforementioned vortices. These often form VDs, which then move inward toward the trap center.


\section{\label{sec:Conclusion}Conclusion}

A nonrotating pancake-shaped Bose--Einstein condensate (BEC) fragmented into three pieces, with a repulsive nonlinearity, forms significant numbers of vortices and antivortices after merging. We have demonstrated, with numerical 2D simulations of this system, that the vortex creation mechanism may be explained in terms of a linear theory of the interference of expanding wave packets.
This was contrasted with the vortex creation mechanism from a two-fragment BEC, in which dark stripe solitons decay into necklaces of vortices and antivortices; an intrinsically nonlinear physical process. With the three pieces separated sufficiently and arranged symmetrically, the formation of a distorted honeycomb lattice containing equal numbers of vortices and antivortices was demonstrated and explained via the linear theory. Moreover, this theory shows that phase differences between the initially separated pieces manifest as a global lattice translation. If allowed to expand in the absence of a trapping potential, the honeycomb lattice maintains its form as it expands. If instead the BEC evolves within a trap, the lattice was shown to melt, exhibiting a diversity of vortex interactions, including the formation of vortex--antivortex dipoles and other vortex clusters.

\acknowledgments{The authors thank S. Clarke, L.D. Turner and A.M. Martin for helpful discussions.
G.R. is supported by an Australian Postgraduate Award. D.M.P. acknowledges
support from the Australian Research Council.}



\appendix
\section{\label{sec:VortexLatticeProduction}Vortex Lattice Production}

In this Appendix we provide an analytical description of the vortices generated through linear superposition of three wavefunction fragments expanding from concentrated Gaussian distributions. The exposition here follows a similar approach to that reported in \citep{RubPa07} which studied the vortices in the interference pattern from the Young's three-pinhole interferometer. In that work, an analytical description of the far-field vortex locations was derived as a function of source arrangement for three equal-amplitude complex scalar waves represented as either spherical waves or pinhole secondary sources. A representation in terms of a discrete parameter space arose, allowing estimates of the number of vortices---shown to relate to the information capacity of a beam---and the description of a natural coordinate system in terms of a family of hyperbolas.

\begin{figure}[bt!]
    \centering
    \includegraphics[width=85mm]{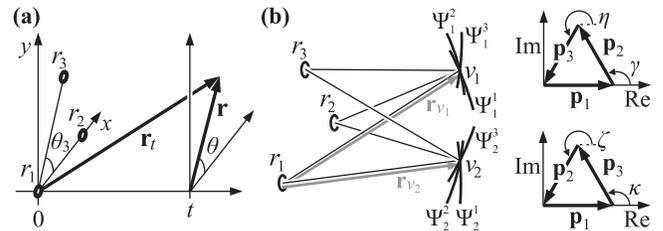}
  \caption{(a) Coordinate system. Three Gaussians lie in the $xy$-plane, centered at points $\gvec{r_1}$, $\gvec{r_2}$ and $\gvec{r_3}$. Vortex positions are parameterized by polar coordinates $(r, \theta)$ at time $t$ where $\gvec{r}_t$ is a (2+1)D position vector. (b) Phasor construction of three interfering waves. Two possible phasor orderings, corresponding to a vortex and antivortex, may be produced by three given wavefunctions $\Psi^s_v, s\in\{r_1,r_2,r_3\}, v\in\{v_1,v_2\}$.}
  \label{fig:pinholes}
\end{figure}

Consider three (2+1)D Gaussian wave packets, each of unit mass, whose position expectation values are stationary and centered about coordinates $\gvec{r_1} \equiv \gvec{0}, \gvec{r_2} ,\gvec{r_3}$. This is shown in the polar coordinate system schematic, Fig.~\ref{fig:pinholes}(a).
The normalized probability amplitude for a Gaussian centered at the origin, as a function of position $\gvec{r}$ and time $t$ is given by \citep{BraJo00_p65}
\begin{equation}
  \label{eqn:2Dgaussian}
    \Psi(\gvec{r},t) = \dfrac{\pi^{-1/2} \Delta p/\hbar}{1+i(\Delta p)^2 t/(m\hbar)} \exp\left( \dfrac{- \left(\Delta p/\hbar \right)^2 |\gvec{r}|^2}
                            {2[1+i(\Delta p)^2 t/(m\hbar)]} \right),
\end{equation}
where $m$ is the mass of an atom of the atomic species, $\Delta p$ is the momentum uncertainty that defines the wave packet width, and $\gvec{r}_t=(\gvec{r},t)$ is a (2+1)D position vector covering the $xy$-plane, whose origin lies at $(\gvec{r},t)=(\gvec{0},0)$.
The total probability amplitude arising from the three superposed Gaussians is then
\begin{multline}
  \label{eqn:general_sum}
    \Psi(\gvec{r},t) = \sum_{j=1}^3 \dfrac{\pi^{-1/2} \Delta p/\hbar}{1+i(\Delta p)^2 t/(m\hbar)}\\
    \times \exp\left(\dfrac{-\left(\Delta p/\hbar \right)^2 |\gvec{r}-\gvec{r}_j|^2}
                           {2[1+i(\Delta p)^2 t/(m\hbar)]} + i\phi_j \right),
\end{multline}
where $|\gvec{r}-\gvec{r}_j| \equiv \sqrt{(x-x_j)^2 + (y-y_j)^2}$ is the distance from the center of the $j$th Gaussian to a given observation point $(\gvec{r},t)$ and $\phi_j$ is the relative phase of the $j$th Gaussian. Separating the amplitude and phase terms of the wave field components $\Psi_j(\gvec{r},t) = A_j(\gvec{r},t) \exp[i\chi_j(\gvec{r},t)]$, we get
\begin{equation}
  \label{eqn:amplitude}
    A_j(\gvec{r},t) =
        \dfrac{\pi^{-1/2} \Delta p/\hbar}{1+i(\Delta p)^2 t/(m\hbar)}
        \exp \left(\dfrac{-(\Delta p)^2 m^2 |\gvec{r}-\gvec{r}_j|^2}
                           {2[(\Delta p)^4 t^2+m^2 \hbar^2]} \right),
\end{equation}
and
\begin{equation}
  \label{eqn:phase}
    \chi_j(\gvec{r},t) = \dfrac{(\Delta p)^4 m t |\gvec{r}-\gvec{r}_j|^2}
                     {2[(\Delta p)^4 t^2\hbar+m^2 \hbar^3]}+\phi_j.
\end{equation}

Vortices lie at points of the phase $\chi(\gvec{r},t)$, at which a line integral of $\nabla_\perp \chi$ along a closed path $\Gamma$ about such a point evaluates to a nonzero value. More formally $\oint_{\ \Gamma} d\chi = 2\pi n$ for some integer $n\neq0$. At any point coinciding with a vortex core, the phasors $\gvec{p_1}, \gvec{p_2}, \gvec{p_3}$ corresponding to the three wave packets, must sum to zero. Equivalently, Eq.~\eqref{eqn:general_sum} is equated to zero. With reference to Fig.~\ref{fig:pinholes}(b), if the phasors are of equal length---corresponding to equal amplitude contributions from the three Gaussians---an equilateral triangle is formed, allowing the angles at the vertices to be specified simply.

As the amplitude contributions from the Gaussians are not unconditionally equal we must make an appropriate approximation to the amplitude term. By restricting consideration to some finite region of the $xy$-plane defined by $|\gvec{r}-\gvec{r}_j| \le |\gvec{r}_\mathrm{max}-\gvec{r}_j|$,
the exponential term in Eq.~\eqref{eqn:amplitude} will be approximately unity provided its argument is small, or $(\Delta p)^4 t^2+m^2 \hbar^2 \gg (\Delta p)^2 m^2 |\gvec{r}_\mathrm{max}-\gvec{r}_j|^2 /2$.
For a given $\gvec{r}_\mathrm{max}$, this is always true after sufficient time has elapsed.
The amplitude term may now be factored out of the probability amplitude expression.
Expanding the term $|\gvec{r}-\gvec{r}_j|^2 = r^2 - 2\gvec{r}\cdot\gvec{r}_j + \gvec{r}_j\cdot\gvec{r}_j$ and remembering that $\gvec{r}_1=\gvec{0}$, the probability amplitude $\Psi(\gvec{r},t)$ from Eq.~\eqref{eqn:general_sum} vanishes when
\begin{equation}
  \begin{split}
    \label{eqn:spherical_phasor}
    1 &+ \exp\left\{i\left[ \alpha \left(r_2^2 - 2 r_2 r \cos \theta \right) + \phi_2 \right]\right\} \\
          &+ \exp\left\{i\left[ \alpha \left(r_3^2 - 2 r_3 r \cos (\theta-\theta_3) \right) + \phi_3 \right]\right\}=0,
  \end{split}
\end{equation}
where $\alpha = m \hbar t / [2(\hbar t)^2+2m^2(\hbar/\Delta p)^4]$. The three summands are associated with the three phasors in Fig.~\ref{fig:pinholes}(b). Whereas the first summand, $1$, is uniquely identified with the horizontal phasor  $\gvec{p_1}$, the association of the sources with the other two phasors allows two permutations, corresponding to the two phasor diagrams; one for each vortex, $v_1$ and $v_2$. In fact one will be a vortex and the other an antivortex.

The arguments of the two exponentials in Eq.~\eqref{eqn:spherical_phasor} are denoted by
$\gamma$ and $\eta$, respectively. These phase angles are uniquely defined to within an integer multiple of $2\pi$, so that:
\begin{subequations}
	\label{eqn:gamma_eta}
    \begin{equation}
        \label{eqn:angle_gamma}
        \gamma = \alpha \left(r_2^2 - 2 r_2 r \cos \theta \right) + \phi_2 = \dfrac{2\pi}{3} + 2m\pi,
    \end{equation}
    \begin{equation}
      \label{eqn:angle_eta}
        \eta = \alpha \left(r_3^2 - 2 r_3 r \cos (\theta-\theta_3) \right) + \phi_3 = \dfrac{4\pi}{3} + 2n\pi,
    \end{equation}
\end{subequations}
where $m$ and $n$ are integers. The association of the $m$ and $n$ indices with the vertices is arbitrary. The choice is made here to match $m$ with $r_2$ and $n$ with $r_3$. The alternative phasor association is thus
\begin{subequations}
	\label{eqn:zeta_kappa}
    \begin{equation}
        \label{eqn:angle_zeta}
        \zeta = \alpha \left(r_2^2 - 2 r_2 r \cos \theta \right) + \phi_2 = \dfrac{4\pi}{3} + 2m\pi,
    \end{equation}
    \begin{equation}
      \label{eqn:angle_kappa}
        \kappa = \alpha \left(r_3^2 - 2 r_3 r \cos (\theta-\theta_3) \right) + \phi_3 = \dfrac{2\pi}{3} + 2n\pi.
    \end{equation}
\end{subequations}
Forming the fraction $\gamma / \eta$ or $\zeta / \kappa$ yields
	\begin{equation}
  \label{eqn:perturbation_start}
    	\dfrac{r_3}{r_2}\left(\vphantom{\dfrac{r_3}{r_2}}\cos \theta_3 +
    	    \tan \theta \sin \theta_3\right) = \dfrac{r_3^2-\beta N(n)}{r_2^2-\beta M(m)},
	\end{equation}
where $\beta = 1/(3\alpha) = 2[(\hbar t)^2+m^2(\hbar/\Delta p)^4]/(3m \hbar t)$.
For the fraction $\gamma / \eta$, we get
\begin{subequations}
    \begin{equation}
    \begin{split}
		\label{eqn:MN}
		M(m) &= 2\pi \left[1+3\left(m - \phi_2/{2\pi} \right)\right], \\
		N(n) &= 2\pi \left[2+3\left(n - \phi_3/{2\pi} \right)\right],
    \end{split}
    \end{equation}
and for the fraction $\zeta / \kappa$, we instead have
    \begin{equation}
    \begin{split}
    \label{eqn:NM}
		M(m) &= 2\pi \left[2+3\left(m - \phi_2/{2\pi} \right)\right], \\
		N(n) &= 2\pi \left[1+3\left(n - \phi_3/{2\pi} \right)\right].
    \end{split}
    \end{equation}
\end{subequations}
Examination of these expressions reveals that a $2\pi$ change in relative phases $\phi_2$ or $\phi_3$ may be absorbed as an integer change in the associated parameter-space coordinate $m$ or $n$, respectively. By allowing $m$ and $n$ to correspond to discrete real values instead of integer values, they may then also absorb fractional parts of the relative phase. Consequently, the vortex lattice may be continuously translated by a single lattice cell for each $2\pi$ change in the source phase. The position of any vortex in a cell coinciding with the BEC trap center is thus understood as resulting from a translation of the entire lattice.
Isolating $\theta$ in Eq.~\eqref{eqn:perturbation_start} yields
\begin{equation}
\label{eqn:theta_mn}
  \theta =
	    \arctan\left[ \dfrac{1}{\sin \theta_3} \left( \dfrac{r_3 - \beta N(n)/r_3}{r_2 - \beta M(m)/r_2} \right) - \cos \theta_3 \right].
\end{equation}
Finally, the expression for the radial coordinate $r$ of the $(m,n)$th vortex core is obtained from Eq.~\eqref{eqn:angle_gamma} or Eq.~\eqref{eqn:angle_zeta} by applying the identity $\cos^2 \theta = 1/(1+\tan^2 \theta)$ and making use of Eq.~\eqref{eqn:theta_mn}:
\begin{equation}
  \label{eqn:r_mn}
    r \!= \!
     \dfrac{1}{2} \sqrt{\dfrac{1}{\sin^2 \theta} \left[r_3 - \dfrac{\beta N(n)}{r_3} - \left(r_2 - \dfrac{\beta M(m)}{r_2}\right) \cos \theta_3\right]^2+1}.
\end{equation}
Together with Eq.~\eqref{eqn:theta_mn}, we have the vortex coordinates $(r,\theta,t)$, with the sign of the vortex charge---i.e. whether they describe the positions of vortices or antivortices---depending on the choice of Eqs.~\eqref{eqn:MN} or \eqref{eqn:NM} and the value of $\theta_3$. If $\theta_3 \in (0, \pi)$, vortices are indicated by Eq.~\eqref{eqn:MN} and antivortices by Eq.~\eqref{eqn:NM}. If $\theta_3 \in (\pi, 2\pi)$, the association is reversed, with vortices indicated by Eq.~\eqref{eqn:NM} and antivortices by Eq.~\eqref{eqn:MN}.

Note that, in contrast with the case described in \citep{RubPa07}, in which the allowed range of integers $(m,n)$ was restricted, here there are no restrictions and an infinite, uniform vortex--antivortex lattice is generated. However, the amplitude term Eq.~\eqref{eqn:amplitude} applies a Gaussian envelope to the probability density $|\Psi|^2 = \left| \sum_{j=1}^3 A_j(\gvec{r}) \right|^2$, effectively limiting the lattice. For sources arranged at the three corners of an equilateral triangle, the lattice has a symmetric honeycomb symmetry, with regular hexagonal cells. Changing the angle $\theta_3$, or the side lengths $r_2$ or $r_3$, distorts the cells and the lattice.

\end{document}